\documentclass[aj,twocolumn,nolinenumbers]{aastex62}

\usepackage{placeins}
\usepackage{comment}

\usepackage{amsmath}
\usepackage{upgreek}

\newcommand{\kms}{\ensuremath{\rm km\,s^{-1}}}

\newcommand{\kp}{\ensuremath{K_{p}}}
\newcommand{\vsys}{\ensuremath{v_{sys}}}
\newcommand{\ra}{\ensuremath{R_{A}}}
\newcommand{\rb}{\ensuremath{R_{B}}}
\newcommand{\Msol}{\ensuremath{\rm M_{\odot}\,}}
\newcommand{\Mj}{\ensuremath{\rm M_{J}\,}}
\newcommand{\Rstar}{\ensuremath{\rm R_{\star}\,}}
\newcommand{\Rst}{\ensuremath{\rm \,R_{\star}}}

\submitjournal{AJ}

\shorttitle{The Atmosphere of KELT-9b with TRES}
\shortauthors{Lowson et al.}

\begin{document}

\title{Multi-epoch detections of the extended atmosphere and transmission spectra of KELT-9b with a 1.5\,m telescope}

\correspondingauthor{Nataliea Lowson}
\email{nataliea.lowson@usq.edu.au}

\author[0000-0001-6508-5736]{Nataliea Lowson} 
\affil{Centre for Astrophysics, University of Southern Queensland, 499-565 West Street, Toowoomba, QLD 4350, Australia}

\author[0000-0002-4891-3517]{George Zhou} 
\affil{Centre for Astrophysics, University of Southern Queensland, 499-565 West Street, Toowoomba, QLD 4350, Australia}

\author[0000-0001-7294-5386]{Duncan J. Wright} 
\affil{Centre for Astrophysics, University of Southern Queensland, 499-565 West Street, Toowoomba, QLD 4350, Australia}

\author[0000-0003-0918-7484]{Chelsea X. Huang} 
\affil{Centre for Astrophysics, University of Southern Queensland, 499-565 West Street, Toowoomba, QLD 4350, Australia}

\author[0000-0002-6907-4476]{Jo\~{a}o M. Mendon\c{c}a} 
\affil{National Space Institute, Technical University of Denmark, Elektrovej, DK-2800 Kgs. Lyngby, Denmark}

\author[0000-0001-9749-6150]{Samuel H. C. Cabot} 
\affil{Yale University, 52 Hillhouse Avenue, New Haven, CT 06511, USA}

\author{Christa Pudmenzky} 
\affil{Centre for Applied Climate Sciences, University of Southern Queensland, 499-565 West Street, Toowoomba, QLD 4350, Australia}

\author[0000-0001-9957-9304]{Robert A. Wittenmyer} 
\affil{Center for Astrophysics | Harvard \& Smithsonian, 60 Garden St., Cambridge, MA 02138, USA}

\author[0000-0001-9911-7388]{David W. Latham} 
\affil{Center for Astrophysics | Harvard \& Smithsonian, 60 Garden St., Cambridge, MA 02138, USA}

\author[0000-0001-6637-5401]{Allyson Bieryla} 
\affil{Center for Astrophysics | Harvard \& Smithsonian, 60 Garden St., Cambridge, MA 02138, USA}

\author[0000-0002-9789-5474]{Gilbert A. Esquerdo} 
\affil{Center for Astrophysics | Harvard \& Smithsonian, 60 Garden St., Cambridge, MA 02138, USA}

\author{Perry Berlind} 
\affil{Center for Astrophysics | Harvard \& Smithsonian, 60 Garden St., Cambridge, MA 02138, USA}

\author[0000-0002-2830-5661]{Michael L. Calkins}
\affil{Center for Astrophysics | Harvard \& Smithsonian, 60 Garden St., Cambridge, MA 02138, USA}




\begin{abstract}
Irradiated Jovian atmospheres are complex, dynamic, and can undergo temporal variations due to the close proximity of their parent stars. Of the Jovian planets that have been catalogued to date, KELT-9b is the hottest Gas Giant known, with an equilibrium temperature of 4050 K. We probe the temporal variability of transmission spectroscopic signatures from KELT-9b via a set of archival multi-year ground-based transit observations, performed with the TRES facility on the 1.5 m reflector at the Fred Lawrence Whipple Observatory. Our observations confirm past detections of Fe I, Fe II and Mg I over multiple epochs, in addition to excess absorption at H$\upalpha$, which is an indicator for ongoing mass-loss. From our multi-year dataset, the H$\upalpha$ light curve consistently deviates from a standard transit, and follows a `W' shape that is deeper near ingress and egress, and shallower mid-transit. To search for and quantify any seasonal variations that may be present, we parameterise a `cometary tail' model to fit for the H$\upalpha$ transit. We find no detectable variations between the different observed epochs. Though a `cometary tail' describes the H$\upalpha$ flux variations well, we note that such a scenario requires a high density of neutral hydrogen in the $n=2$ excited state far beyond the planetary atmosphere. Other scenarios, such as centre-to-limb variations larger than that expected from 1-D atmosphere models, may also contribute to the observed H$\upalpha$ transit shape. These multi-epoch observations highlight the capabilities of small telescopes to provide temporal monitoring of the dynamics of exoplanet atmospheres.
\end{abstract}



\section{Introduction}

Deciphering an exoplanet atmosphere signal from that of its host star has historically been a difficult endeavour, initially being restricted to either space based telescopes (e.g. \textit{HST} and \textit{Spitzer}) or large aperture ($\geq 8$ m) ground based telescopes \citep[see reviews by][]{2010ARA&A..48..631S,Madhusudhan_2019}. Thanks to advances in our techniques for probing exoplanet atmospheres, and the efforts by ground and space-based surveys for identifying new transiting planets suitable for characterisation~\citep{Knutson_2009,Stevenson_2014,Birkby_2018}, we can now demonstrate that metre-class facilities are also capable of characterising exoplanet atmospheres, as presented in this publication.

Many exoplanet atmosphere observations focus around the primary transit, where we analyse light from the host star that travels through the day-night terminator region of the planet as it transits in front of its host star. One of the observational techniques that captures these transits is known as the high resolution spectroscopy technique. This involves using a high resolution spectrograph, $R\geq25,000$, to detect the Doppler shift of the planet as it travels along its orbit over the course of the transit. For close-in planets, this is achievable due to the motion of the planet moving at a larger velocity in comparison to the simultaneously observed stellar spectra of the host star, varying as per the barycentric velocity shift, and the near-stationary telluric absorption features~\citep[see review by][]{Birkby_2018}. Chemical composition is determined by comparing the detected spectral lines to high resolution spectra generated from modelling codes with the same physical parameters (such as temperature) that is calculated for the observed planet. In addition to chemical composition, this technique can decipher additional physical parameters of an exoplanet, including true planetary mass~\citep[e.g.][]{deKok_2013}, temperature profiles~\citep[e.g.][]{Snellen_2010}, the presence of clouds (when optical spectra is used in conjunction with low resolution near infrared transmission spectra)~\citep[e.g.][]{Zak_2019,Allart_2020}, day-to-night winds~\citep[e.g.][]{Snellen_2010,Louden_2015}, and the rotation period of the planet~\citep[e.g.][]{Brogi2016}.

Analysing the chemical composition of an exoplanet atmosphere through high resolution spectroscopy not only tells us what atoms are in the atmosphere, but can reveal which ones are escaping it. For exoplanets that reside close to their host stars, the exposure to extreme levels of irradiation enables active atmospheric evaporation from the planet ~\citep[e.g.][]{Lammer_2003, Yelle_2004, Garcia_Munoz_2007, Murray-Clay_2009, Owen_2012, 2021_King, 2022_Kubyshkina}. Mass loss through evaporation of a primordial atmosphere is one of the dominant evolutionary drivers for atmospheres, often occurring within the first billion years after formation~\citep[e.g.][]{Owen_2019, Howe_2020, Bean_2021}.

Atmospheric escape can be identified via absorption lines by atoms with a planetary radius beyond the Roche limit (a region where the gravity of the planet equals that of the parent star). Absorption by species close to the Roche limit, and at high velocities with respect to the planet, can also be tracers of ongoing escape. The most successful observational tracer for atmosphere evaporation involves searching for excess Lyman-$\upalpha$ (Ly$\upalpha$)~\citep[e.g.][]{Vidal-Madjar_2003, Lecavelier_2010, Ehrenreich_2015, Bourrier_2018, 2020_Odert}, while He I 10830 \AA{} has also been identified as invaluable~\citep[e.g.][]{Spake_2018, Allart_2018, Mansfield_2018, Salz_2018, Alonso-Floriano_2019, Ninan_2020, Kirk2020, Kirk2022}. However, with the Ly$\upalpha$ absorption line occurring in UV (1215.67 \AA{}) and He I 10830 \AA{} occurring in the near infrared, detection of these species are mostly limited to space-based telescopes or large ground-based telescopes. Fortunately, the optical transmission absorption line of Hydrogen-$\upalpha$ (H$\upalpha$), has also been identified as a potential indirect probe for ongoing atmosphere evaporation, enabling this process to be observed with ground based high resolution spectroscopic facilities~\citep[e.g.][]{Cauley_2017, Casasayas-Barris_2018, Cabot_2020, Yan_2021, Czesla_2022}. Observing absorption lines alluding to planetary mass loss will help resolve exoplanet evolutionary enigmas, such as the Hot Neptune `desert' and the radius-valley, both of which have been proposed to be the product of atmosphere evaporation during the late stages of planet formation~\citep[e.g.][]{Beauge_2013, Mazeh_2016, Lundkvist_2016, Fulton_2017, Fulton_2018, Van_Eylen_2018, 2020_Venturini, 2021_Rogers}.

One of the notable exoplanets where ongoing detections of H$\upalpha$ have been observed in its upper atmosphere is the Ultra Hot Jupiter KELT-9b~\citep{Yan_2018}. With $T_{eq}=4050\pm180$ K, KELT-9b is the hottest Jovian exoplanet discovered thus far, having a dayside temperature ($\sim$4900 K) equivalent to the photospheric temperature of K-stars~\citep{Gaudi_2017, Hooton_2018}. A number of chemical elements have been detected in both primary and secondary transit observations of KELT-9b~\citep{Cauley2019, Hoeijmakers_2019, Turner_2020, Yan_2019, Pino_2020, 2021_Changeat}, including Fe II, which had never been observed in an exoplanet atmosphere prior to KELT-9b~\citep{2018_Hoeijmakers, 2022_Bello-Arufe}. Detections of excess H$\upalpha$ absorption in the atmosphere of KELT-9b have been repeatedly linked to atmosphere evaporation~\citep{Yan_2018,  Cauley2019, 2020_Wyttenbach}, which has been proposed to be the product of thermal dissociation and recombination of H$_2$ in the upper atmosphere of the planet in the presence of strong UV irradiation~\citep{Yan_2018, Kitzmann_2018, 2019_Munoz, 2020_Mansfield}.~\citet{Cauley2019} pointed out significant substructure in the absorption time series of metal and Balmer lines for their 2019 transit using the PEPSI instrument on the Large Binocular Telescope. They specially noted the blueshifted absorption extending to $\sim$100 \kms{} in the line profile during egress, and hypothesised it to be due to a wind-like geometry where material is being accelerated away from the planet towards the observer. They proposed this wind is due to a temporal spike in the planet's mass loss rate caused by a stellar flare. Such phenomena were not present in the two H$\upalpha$ absorption time series observed in \citet{Yan_2018} using CARMENES, however the signal-to-noise in their dataset prohibited them to resolve and measure velocity centroids of individual transmission spectra. 

In this paper, we use two epochs of archival observations of KELT-9b taken on a 1.5 m telescope to characterise its atmosphere. We also debut a geometric model to monitor temporal variability between epochs, which is derived using the `W' profile displayed in the H$\upalpha$ transit photometry. In Section~\ref{sec:obs}, we outline how the observations were taken and our method for telluric subtraction. Section~\ref{sec:h-alpha_analysis} details the extraction of the H$\upalpha$ absorption, construction of our geometric model, and comparing it against our photometric data sets. Section~\ref{sec:atmos_cc} presents our additional detections of metals and our mass estimations for KELT-9 and KELT-9b, followed by our discussion and conclusion in Section~\ref{sec:results} and~\ref{sec:summary} respectively.

\section{Observations and Telluric Subtraction}\label{sec:obs}
We used archival observations available for KELT-9 from the Tillinghast Reflector Echelle Spectrograph (TRES) on the 1.5\,m reflector at the Fred Lawrence Whipple Observatory, FLWO (Mount Hopkins, Arizona, USA). TRES is a fiber-fed echelle with a resolving power of $\uplambda / \Delta \uplambda \equiv R=44,000$, covering the spectral range of $3850$\,\AA{}$-9100$\,\AA{} over 51 echelle orders~\citep{TRES}.

These archival observations are the same data set presented in the KELT-9b discovery paper~\citep{Gaudi_2017}, with a total of 75 spectra being observed over three separate transit epochs (UT: 2014 November 15, 2015 November 06, 2016 June 12). Observations on 2014 November 15 were obtained at an exposure time of 720\,s, achieving $\mathrm{S/N}\approx300$ per resolution element over the Mg b lines. Observations on 2015 November 06 had an exposure time of 540\,s, achieving $\mathrm{S/N}\approx140$. Observations on 2016 June 12 yielded far lower S/N spectra, and were subsequently not used in the remainder of this analysis. 

Ground-based observations are always contaminated by telluric absorption through the Earth's atmosphere. These absorbing species within Earth's atmosphere (H$_{2}$O and Na in particular) interact with the incoming light from the host star prior to reaching our detectors. These features contaminate the minute absorption signatures we are attempting to retrieve. We follow the general techniques adopted by similar previous analyses~\citep[e.g.][]{2019_Cabot}, and remove these telluric features via a set of synthetic models. In this paper, we make use of the \texttt{telfit} module \citep{2014AJ....148...53G} to model the atmosphere via the Line-By-Line Radiative Transfer Model \citep{Clough:1992}. 

\begin{figure}
    \begin{center}
    \includegraphics[width=\columnwidth]{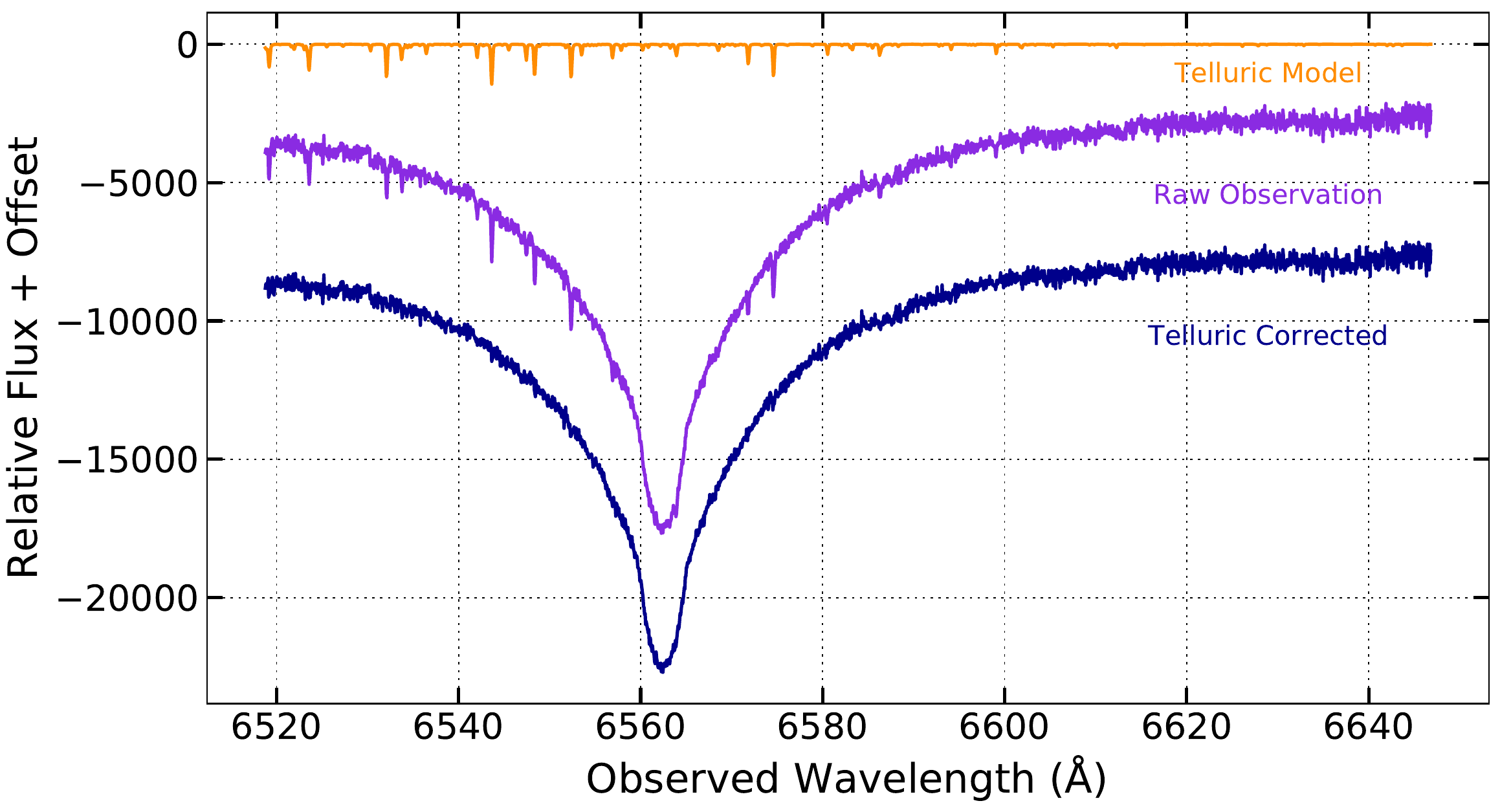}
    \caption{Example telluric correction over the H$\upalpha$ wavelengths for the 2014 KELT-9b observations. We generate a library of synthetic telluric spectra using the \texttt{telfit} implementation of a line-by-line radiative transfer model. The observations are then matched against the interpolated library. The resulting corrected spectrum is shown in the bottom.}
    \label{fig:telluric}
    \end{center}
\end{figure}

To fit the observed telluric lines, we produced a set of $\sim$10,000 models to sufficiently explore a parameter space varying for humidity, oxygen mixing ratio, zenith angle, and instrument resolution. This library is interpolated using a gradient boosting regressor via the \texttt{scikit-learn} package~\citep{scikit}. The best fitting telluric model is identified via a least-squares fit between the observations and the telluric library. Despite the telluric corrections, we still discard spectral orders severely influenced by the telluric O$_2$ absorption bands at $7534$\,\AA{}$-7682$\,\AA{}, and $8922$\,\AA{}$-9097$\,\AA{}. A portion of the telluric-corrected spectrum is presented in Figure~\ref{fig:telluric}.

\section{An extended H$\upalpha$ atmosphere}\label{sec:h-alpha_analysis}
Highly irradiated gas giants are expected to undergo atmospheric escape throughout their lifetimes \citep[see review by][]{Owen_2019}. Excess absorption in Ly$\upalpha$ has been a predominant tracer for atmosphere escape~\citep[e.g.][]{2003Natur.422..143V}, with ``photoevaporation" and core-powered mass-loss being the widely accepted models used to explain this process~\citep{2021_Owen}. Likewise for optical band observations, excess absorption of H$\upalpha$ is being increasingly reported near the Roche radius for numerous highly irradiated hot Jupiters \citep{Casasayas-Barris_2019, Chen_2020, Cauley_2021,Yan_2021, Czesla_2022}, including KELT-9b \citep{Yan_2018,Cauley2019,2020_Wyttenbach}. While the detection of H$\upalpha$ in KELT-9b has been interpreted as a signature of active mass loss by some~\citep[e.g.][]{Yan_2018, 2019_Munoz}, it has been contested by others~\citep[e.g.][]{2020_Turner, 2020_Fossati}.

PEPSI observations in 2018 by \citet{Cauley2019} showed that KELT-9b exhibits a strong H$\upalpha$ absorption in-transit, with a `W' shaped transit light curve. We describe below our efforts to recover and model this effect in our observations from 2014 and 2015.

Section~\ref{sec:lc_extract} presents an overview on the removal of the stellar spectrum, the white light Doppler tomographic planetary transit, and recovery of the planetary H$\upalpha$ excess signal through the transit event. Section~\ref{sec:halpha_lc} presents a toy transit model to describe the H$\upalpha$ transit light curves from our observations.

\subsection{H$\alpha$ transit light curves}\label{sec:lc_extract}
In this section, we detail the analysis of the H$\upalpha$ excess absorption about KELT-9b from our TRES observations. We detected the excess H$\upalpha$ absorption of KELT-9b at the expected orbital velocity of the planet during both TRES transit observations. The H$\upalpha$ transit light curve, which maps the temporal variation of the H$\upalpha$ excess through the transit, does not follow the shape expected for a standard white light transit. We discuss our interpretation of this signal, and offer a simple `cometary tail' model that replicates the observed transit shape.

We first normalise the spectral region within 200 \kms{} of the 6562.8 \AA{} H$\upalpha$ absorption feature for each TRES observation. The planetary and stellar signals dramatically differ in their velocity variation over the course of the transit, allowing us to differentiate between the two signals despite their contrast. The planetary transmission signal is expected to vary over $\sim$130 \kms{} during the course of the transit due to its orbital motion, while the stellar Doppler motion is only expected at the $\sim$0.25 \kms{} level. We generate a master spectral template from a median combination of the observed stellar spectra for a given night of observations, and remove this from each observed TRES spectrum via division.

During a transit, the residual spectral signatures comprise of contributions from the Doppler shadow (the Rossiter-McLaughlin effect \citealt{1924_Rossiter, 1924_McLaughlin}) and the transmission spectroscopic signatures from the atmosphere of the planet. Depending on the projected orbital obliquity and the velocity amplitude of the planet's orbit, there is a region in the transit where the Doppler and transmission signals can overlap and cancel each other's effects. When this occurs, the Doppler shadow manifests as a reduction in the apparent absorption in a spectral line, while the transmission signal manifests as an excess of absorption. To correct for this, we simultaneously model for Doppler shadow and planetary transmission spectrum.

\begin{figure}
    \centering
    \includegraphics[width=0.8\columnwidth]{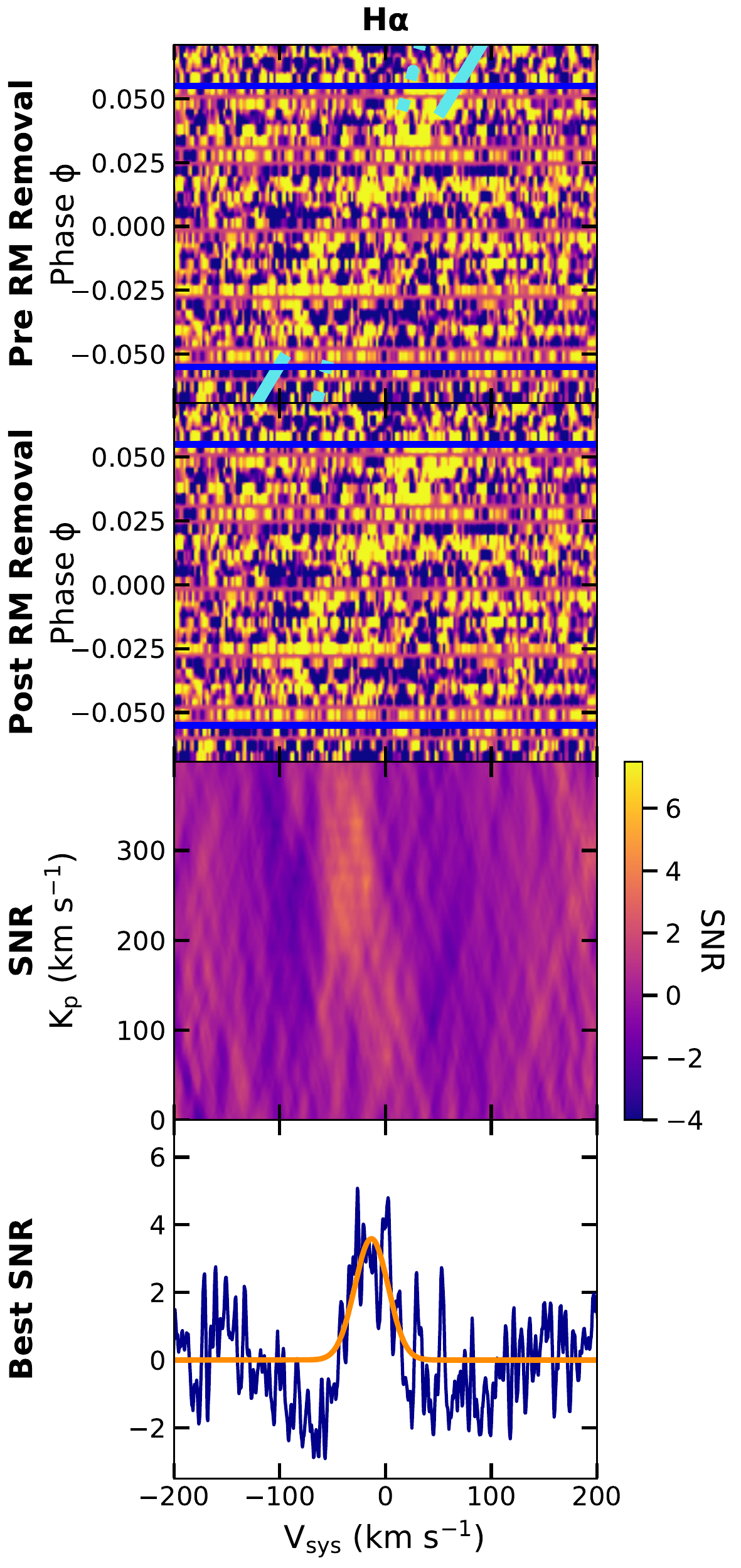}
    \caption{Cross correlation between the ensemble of observations against the H$\upalpha$ absorption of KELT-9b. \textbf{Top:} the cross correlation function analysis without the Doppler shadow subtracted (the solid cyan line represents the trail of the planet and the dashed cyan line represents its Doppler shadow), \textbf{1$^\text{st}$ middle:} the cross correlation function analysis with the Doppler shadow subtracted, \textbf{2$^\text{nd}$ middle:} the cross correlation signal-to-noise of as a function of the planet's orbital velocity \kp{} and systemic velocity \vsys{} \textbf{bottom:} the cross correlation function at the best fit orbital velocity.}
    \label{fig:h-alpha}
\end{figure}

The trail of the Doppler shadow is modelled as per~\citet{2016MNRAS.460.3376Z}, with the transit parameters describing the transit centroid $t_0$, period $P$, normalised semi-major axis $a/$\Rst, inclination $i$, and radius ratio $r_p/\Rst$ fixed to that reported in \citet{Gaudi_2017}. From observing the secondary eclipse phase of KELT-9b, \citet{Wong_2020} measured an eccentricity of $e<0.007$ to 2$\upsigma$ significance, therefore, the planet transmission spectrum is assumed to have the orbital velocity of a circular orbit. We perform a cross correlation between the spectral residuals and a synthetic planetary spectrum, as is appropriate to reveal the planetary transmission spectrum. The Doppler transit signal is best revealed when the template best matches that of the host star spectrum, and the relative cross correlation function height between the planetary signal and the Doppler signal changes based on the specific synthetic planetary spectrum, and the specific species adopted for a given analysis. Therefore, to best remove the Doppler shadow signal, we scale the relative depths of the Doppler shadow and the transmission spectrum trail in our simultaneous fit. The Doppler shadow is then subtracted from our spectral data, leaving only the contribution from the planet's atmosphere trail. The H$\upalpha$ transmission and Doppler shadow transit signals before and after subtraction are presented in Figure~\ref{fig:h-alpha} via the first and second panels.

To estimate the detection significance of the H$\upalpha$ transmission signal, we compute its signal-to-noise ratio over an array of possible orbital velocities for the planet. We perform a grid search over the systematic velocity of the KELT-9 system, \vsys{}, and the radial velocity semi-amplitude of the planet, \kp{}. At each grid point, we align the H$\upalpha$ residuals as per a circular orbit, and average over all exposures captured between second and third contact (i.e. full transit). Figure~\ref{fig:h-alpha} shows the signal-to-noise ratio of the cross correlation peak as a function of \kp{}, and \vsys{} (2$^{\text{nd}}$ middle panel), as well as the cross correlation function signal-to-noise ratio at the predicted \vsys{} of the system (bottom panel). We find the H$\upalpha$ transmission signal can be best traced by a circular orbit with \kp{} $=260\pm110$ \kms{} and \vsys{} $=-24\pm18$ \kms{}.

The H$\upalpha$ transit light curve describes the strength of the H$\upalpha$ absorption during a transit observation. To determine the H$\upalpha$ flux for a given observation, we model the planetary H$\upalpha$ line profile as a Gaussian, with its integral defining its absorption strength.

Due to the intrinsic low signal-to-noise of the planetary absorption feature, we seek to reduce the flexibility of the Gaussian model fit during each epoch. We assume the line profile width does not vary over the course of the observations. We also assume the velocity of the planetary transmission signal follows that of the planetary circular orbit prescribed above. The width $\upsigma$, orbital velocity amplitude \kp{}, and systemic offset \vsys{} are determined from a 2D elliptical Gaussian function fit to the master line profile, constructed from a median combination of all in-transit observations.

To compute the transit light curve, we model the local stellar spectrum blocked by the traversing planet via a Gaussian profile. The Gaussian profile has width $\upsigma$, and follows an orbit with amplitude \kp{}, and systemic \vsys{}. The resulting light curve describing the strength of the planetary H$\upalpha$ absorption through the transits is shown in Figure~\ref{fig:h-alpha_lc}. We note that the per-point uncertainties of the light curve have been scaled such that the reduced $\upchi^2$ of the eventual light curve is at unity, after removal of the best-fit model from Section~\ref{sec:halpha_lc}.

\subsubsection{Estimating the effect of Centre-to-Limb Variations}\label{sec:centretolimb}

The light curves representing transmission spectroscopic signals of deep Fraunhofer lines can be significantly affected by centre-to-limb effects. During a transit, the line profile of the planetary atmospheric spectrum is modulated by the flux of the deep stellar absorption feature. This effect naturally induces a `W' shape to the observed transmission spectrum light curve  \citep[e.g.][]{Snellen2008, 2012MNRAS.426.2483Z, 2015A&A...577A..62W, 2016AJ....152...20C, 2017_Khalafinejad, Cauley2019}.

To model this effect, we first compute the H$\upalpha$ line of KELT-9 at different limb angles corresponding to the phases of each spectral observation. We make use of the \texttt{SPECTRUM} spectral synthesis code \citep{Gray1994} to compute the local H$\upalpha$ line profile as per \citet{Czesla2015}. The planetary absorption feature is modelled as a Gaussian of width $18\,\mathrm{km\,s}^{-1}$ (as per its measured width) at the Keplerian velocity of the planet. The result is a net decrease in the relative absorption of the planet during mid-transit, where the velocity of the planet aligns with that of the systemic velocity of the star. This model is incorporated in the following modelling described in Section~\ref{sec:halpha_lc}. The dotted line in Figure~\ref{fig:h-alpha_lc} represents the best-fit model of the H$\upalpha$ absorption light curve accounting only for this centre-to-limb effect.

\subsection{H$\alpha$ Light Curve Model}\label{sec:halpha_lc}
\citet{Yan_2018} showed that the H$\upalpha$ radius of KELT-9b extends to 70\% that of its Roche lobe, inferring that atmospheric escape may be ongoing for the highly irradiated planet. If the extended atmosphere of KELT-9b is non-spherical, as may be the case due to significant ongoing mass-loss, then the observed transits will be asymmetric.

To examine and quantify any temporal variability in the H$\upalpha$ transit light curve, we present a toy model that parameterises a cometary-tail shaped transit geometry to model the observed `W' shaped  transit. The model illustrates a neutral hydrogen tail being directed radially away from the star, towards the observer. We note however that neutral hydrogen in the $n=2$ excited state is not expected to be present far from the planet atmosphere (Section~\ref{sec:n2}). This toy model nevertheless presents few free parameters, and helps to quantify any variations between the multiple epochs of observations.

\begin{figure}
    \centering
        \includegraphics[width=0.6\columnwidth]{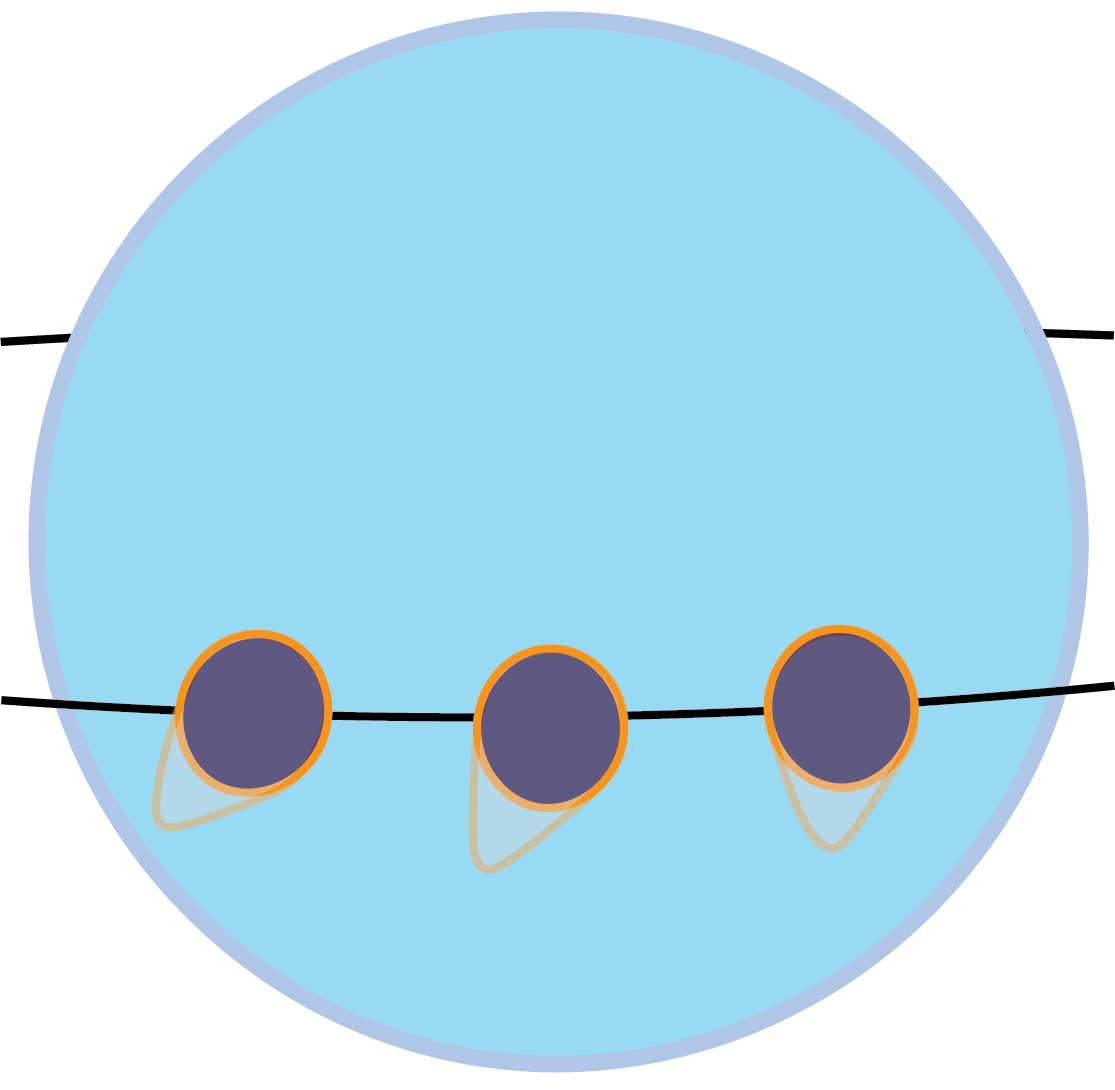}
        \includegraphics[width=0.6\columnwidth]{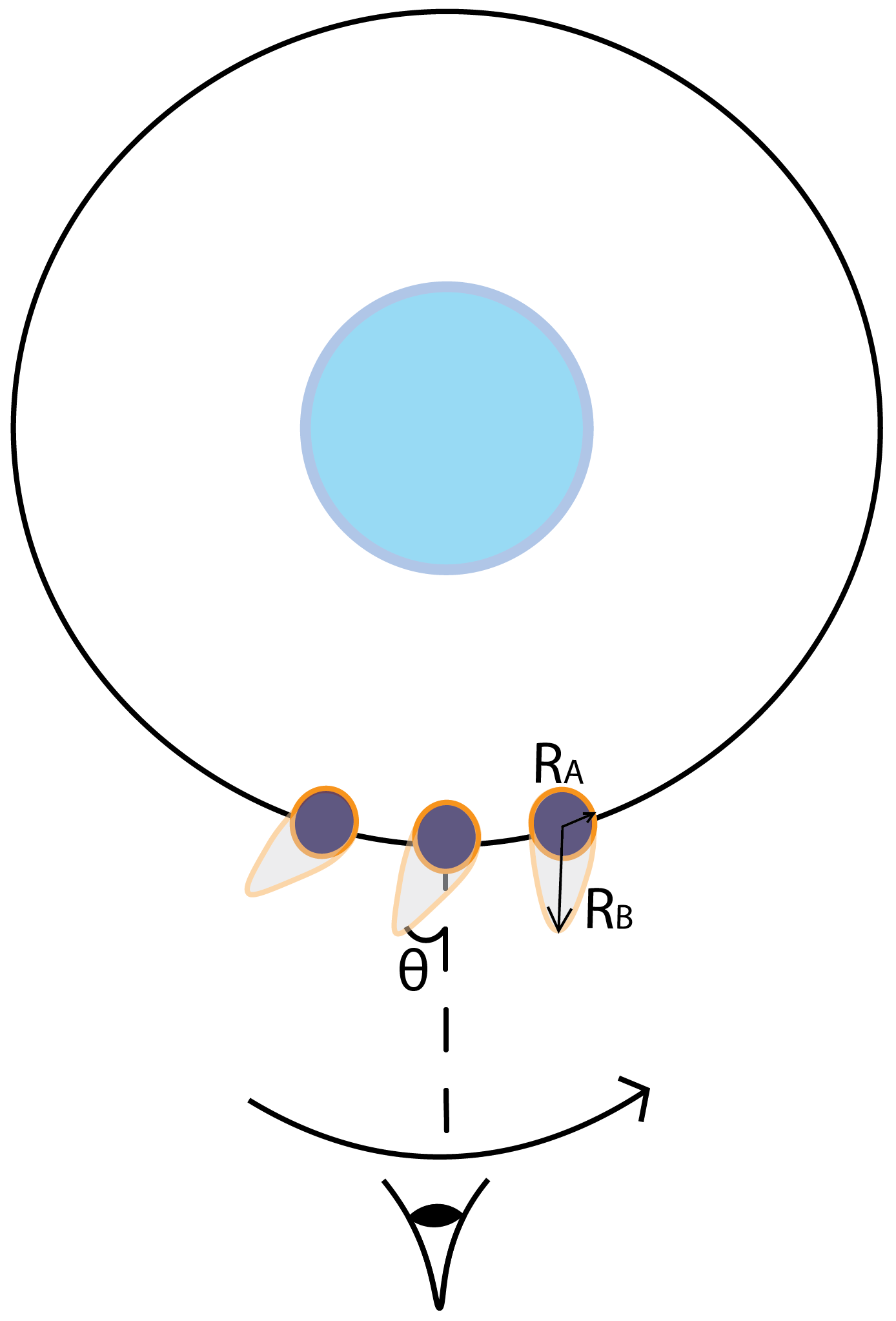}
    \caption{We present one model that successfully reproduces the observed `W' shaped H$\upalpha$ transit of KELT-9b. In this toy model, the planet hosts a comet-like tail pointed away from the star. In this geometry, the area covered by the planet and tail is greatest during ingress and egress, and is reduced during mid-transit, resulting in a `W' shaped transit as per our observations. The figure shows this toy model, with the tail pointed away from the star towards the observer, moving from right to left through the transit. The tail is modelled by a semi-ellipse with a short axis of radius  similar to that of the planet $\ra{} \approx R_p$, and a tail of length \rb{} extending away from the planet towards the observer. Additional asymmetries in the transit are modelled by including a small tilt to the tail ($\uptheta$), trailing away from the direction of motion. With only three free parameters in this toy model, we can easily compare the H$\upalpha$ transits between different observations, and search for temporal variability in the H$\upalpha$ absorption from the planet's atmospheres.}
    \label{fig:geometry}
\end{figure}

In this model, the in-transit absorption is `W' shaped because we see more of the tail during ingress and egress, less when it is aligned with our line of sight. Additional asymmetry in the transit, after inclusion of the centre-to-limb effects described in Section~\ref{sec:centretolimb}, can be explained by a slight tilt in the angle of the tail along the orbital plane, perhaps due to the orbital motion of the planet. We test this hypothesis by constructing a semi-ellipsoid to represent the cometary tail of the planet's escaping atmosphere. As shown in Figure~\ref{fig:geometry}, we denote the X-axis to be along the line-of-sight towards the observer, the Y-axis represents the horizontal axis along the sky plane and the Z-axis is the vertical axis in the sky plane. The combined planet and atmosphere evaporation is represented as a semi-ellipsoid, symmetric along the Y-axis and Z-axis but elongated away from the host star along the X-axis. The elongation is representative of the tail trailing behind the planet. Its projected area on the sky plane can be computed as half of that from the projection of the full ellipsoid except when the elongated axis is aligned with the X-axis exactly.

To compute the projected shape of an ellipsoid on the sky plane conveniently, we use the quadratic form of the ellipsoid to represent it. For example, an elongated ellipsoid along the X-axis can be expressed with equation $\frac{x^2}{a^2} + \frac{y^2}{b^2} + \frac{z^2}{b^2} -1 = 0$, which can have a corresponding matrix of
\begin{equation*}
Q = \begin{pmatrix}
\frac{1}{a^2} & 0 & 0 & 0 \\
0 & \frac{1}{b^2} & 0 & 0  \\
0 & 0 & \frac{1}{b^2} & 0  \\
0 & 0 & 0 & -1 
\end{pmatrix}.
\end{equation*}
We assume the ellipsoid rotates counter-clock-wise around the Z-Axis by an angle $\uptheta$ (due to the movement along the orbit of the planet). The rotated ellipsoid can therefore be expressed with the matrix $Q'=I^{-1}\,Q\,I$, where $I$ represents the rotation matrix along the inclination, $i$, of the planet rotated around the Y-Axis. $i=90^{\circ}$ when the impact parameter of planet is 0.
\begin{equation*}
I = 
\begin{pmatrix}
\sin\,i & 0 & \cos\,i & 0 \\
0 & 1 & 0 & 0 \\
-\cos\,i & 0 & \sin\,i & 0 \\
0 & 0 & 0 & 1 
\end{pmatrix},
\end{equation*}
The projection of $Q'$ onto any plane should be an ellipse represented by the matrix $C = (P(Q')^{-1}P^T)^{-1}$, where $P$ is the camera matrix. The camera matrix $P$ for a sky plane projection on the $Y$ and $Z$ axes is 
\begin{equation*}
P=
\begin{pmatrix}
0 & 1 & 0 & 0 \\
0 & 0 & 1 & 0 \\
0 & 0 & 0 & 1
\end{pmatrix}
.
\end{equation*}
The resulting matrix,
\begin{equation*}
C=
\begin{pmatrix}
\frac{1}{\upbeta^2} & 0 & 0 \\
0 & \frac{1}{\upgamma^2} & 0 \\
0 & 0 & -1 
\end{pmatrix}
,
\end{equation*}
can be interpreted as an ellipse with the familiar function form of $\frac{y^2}{\upbeta^2} + \frac{z^2}{\upgamma^2} = 1$.

We compare our derived H$\upalpha$ light curve to the toy light curve model described above via an MCMC exercise. We account for free parameters describing the semi-ellipsoid radius along the Y-axis and Z-axis (\ra{} / \Rstar), radius along the X-axis (\rb{} / \Rstar), and tilt of the tail ($\uptheta$). Our model also includes the centre-to-limb variation describe in Section~\ref{sec:centretolimb}, with the obliquity of the planet orbit assumed to be $\lambda=-84^\circ$  as per \citet{Gaudi_2017}. Our MCMC uses 50 walkers over 5000 iterations per walker to explore the posterior probability distribution. We analyse the 2014 and 2015 light curves independently, and then model the joint dataset. Free parameters include the radius and tilt parameters \ra{}, \rb{} and $\uptheta$, and two free parameters allowing for a linear786 trend to the light curve. The model also includes the standard transit parameters of normalised semi-major axis $a/$\Rst, line-of-sight inclination $i$, transit epoch $T_0$, and orbital period $P$, the values of which are adopted and fixed to that from \citet{Gaudi_2017}. For our combined 2014 and 2015 analysis, a separate linear trend is allowed for each night of observations. To prevent our walkers from exploring unrealistic values, we restrict our \ra{} and \rb{} parameter spaces to be positive while constraining $\uptheta$ between -90 deg to 90 deg. For each epoch, the best-fit \ra{}, \rb{} and $\uptheta$ parameters are determined from the median of their respective posterior distributions.

\begin{figure}
    \centering
        \includegraphics[width=0.8\columnwidth]{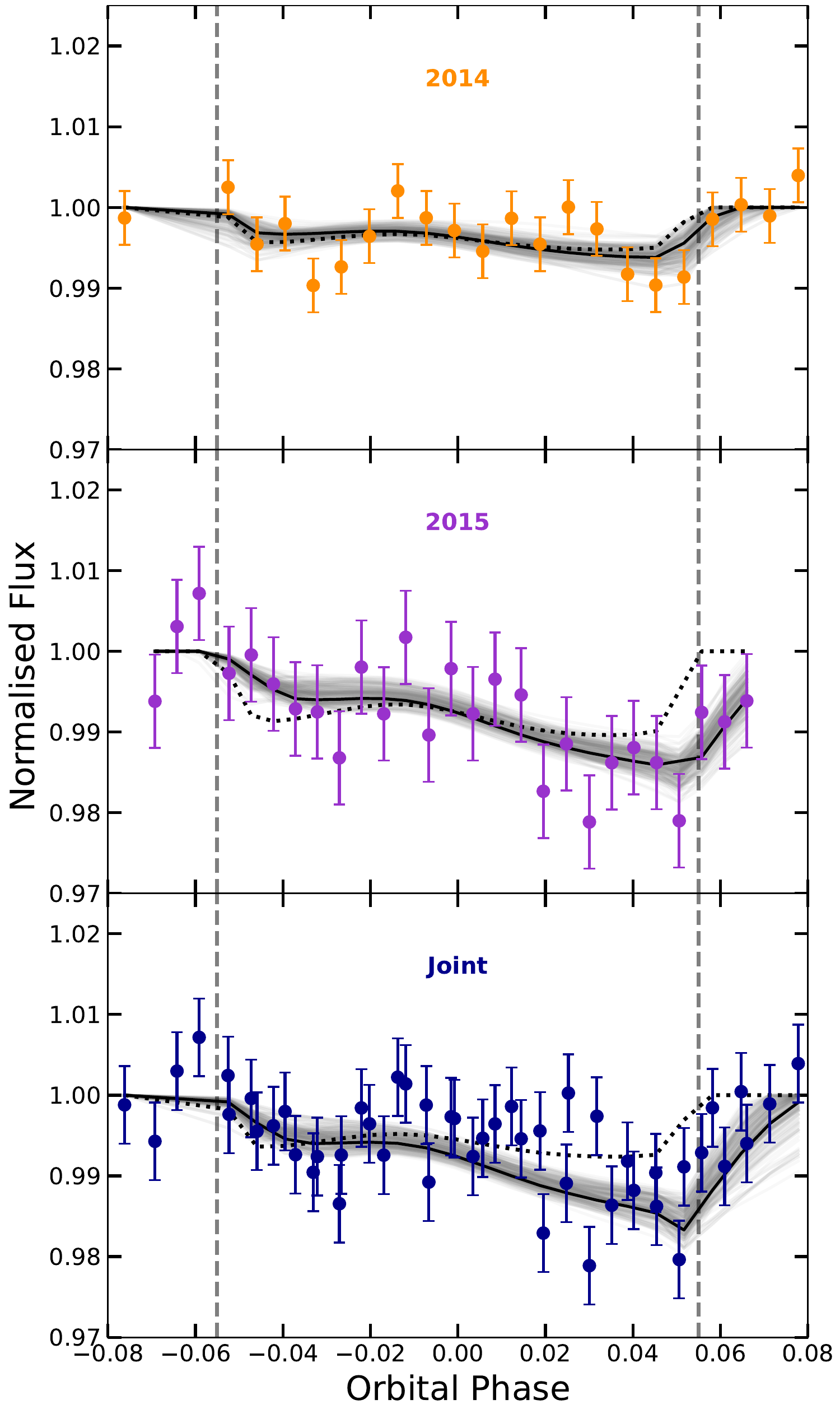}
         \caption{Light curve showing the time variation of the H$\upalpha$ absorption across each transit. The H$\upalpha$ light curve for the the 2014 (\textbf{top}) and 2015 (\textbf{middle}) epochs are plotted independently. The best fit escaping tail model is marked by the solid black line on each panel while the spherical (no tail) model is marked by the dotted line. We also jointly fit the 2014 and 2015 observations (\textbf{bottom}), with the grey lines representing 200 randomly drawn models from our posterior. The vertical dashed lines represent the ingress and egress for each epoch.}
    \label{fig:h-alpha_lc}
\end{figure}

\begin{figure}
    \centering
        \includegraphics[width=0.95\columnwidth]{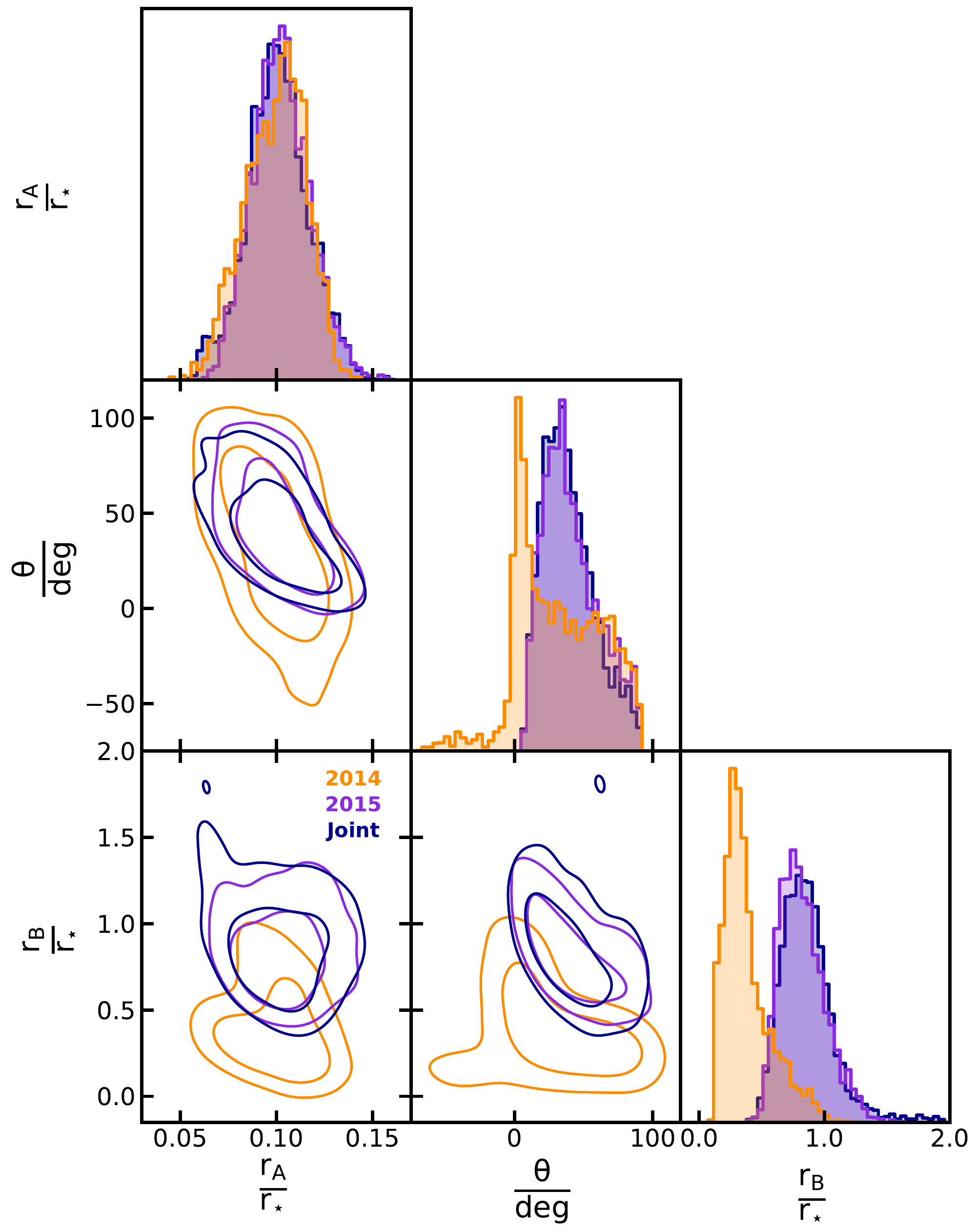}
        \caption{The derived parameters values for \ra{}, \rb{}, and $\uptheta$ when fitting our projected geometric model against the 2014 (\textbf{orange}) and 2015 (\textbf{purple}) epochs, and a combined (\textbf{navy}) data set via MCMC. The inner circle for each contour represents $1\upsigma$ accuracy while the outer represents $2\upsigma$. The histograms display the posterior distribution for each parameter in the 2014, 2015 and joint scenarios.}
    \label{fig:contours}
\end{figure}

Figure~\ref{fig:h-alpha_lc} and Table~\ref{tab:values} present the best fit models for our individual and combined 2014 and 2015 H$\upalpha$ light curves. Figure~\ref{fig:contours} presents the $1\upsigma$ and $2\upsigma$ comparison amongst these free parameters. No significant variability was detected between the two epochs of observations at the $3\sigma$ level.

\begin{deluxetable}{cccc}[h!]
\tablewidth{0pc}
\tabletypesize{\scriptsize}
\tablecaption{
        The best fit values of our H$\upalpha$ transit model
    \label{tab:values}
}
\tablehead{
    &
    & 
    &
     \\
    \multicolumn{1}{c}{Epoch}          &
    \multicolumn{1}{c}{\ra{} (\Rstar)}  \tablenotemark{a}           &
    \multicolumn{1}{c}{\rb{} (\Rstar)} \tablenotemark{b}           &
    \multicolumn{1}{c}{$\uptheta$ (deg)} \tablenotemark{c}            
}
\startdata
\textbf{2014} & $0.101_{-0.018}^{+0.014}$ & $0.33_{-0.12}^{+0.24}$ & $28_{-25}^{+39}$  \\ 
    \textbf{2015} & $0.102_{-0.015}^{+0.015}$ & $0.80_{-0.15}^{+0.20}$ & $38_{-16}^{+27}$\rule{0pt}{4ex} \\ 
    \textbf{Joint} & $0.101_{-0.015}^{+0.016}$ & $0.84_{-0.17}^{+0.20}$ & $36_{-15}^{+22}$\rule{0pt}{4ex} \\ \hline
\enddata 
\tablenotetext{a}{
The length of the short-axis of the half-ellipse, in units of stellar radii.
}
\tablenotetext{b}{
The length of the long-axis of the half-ellipse, in units of stellar radii.
}
\tablenotetext{c}{
Tilt in the ellipse with respect to the orbit normal. 
}
\end{deluxetable}

We caveat that our fit does not incorporate the velocity profile of the outflow. \citet{Owen_2019} modeled the velocity profile of ongoing Ly$\upalpha$ escape, and showed that a net blueshift line profile is expected. Figure~\ref{fig:ha_profile} shows that we measure a broadened H$\upalpha$ velocity profile, as compared to the detected metallic absorption features (Section~\ref{sec:atmos_cc}). A full model that describes the line profiles will include an opacities profile for the escaping gas, and possibly invoke more complex modeling of stellar wind interactions that go beyond the scope of this study.

\section{Searching for metallic absorption features}\label{sec:atmos_cc}

At temperatures of $\sim$4900 K on the day side of KELT-9b, we expect complete disassociation of molecular species commonly found in planetary atmospheres. Instead, we expect the presence of atomic species such as Fe I, Fe II, Mg I, Ca I, Ca II, Cr II, Sc II, Ti II and Y II, all of which have been identified in previous publications performing retrieval analysis on KELT-9b~\citep{2018_Hoeijmakers, Cauley2019, Hoeijmakers_2019, Yan_2019, Pino_2020, Turner_2020, Bello-Arufe_2022}. We use our TRES observations to search for the transmission spectrum from the upper atmosphere of the planet.

First, we remove the stellar spectrum as per Section~\ref{sec:lc_extract}), via the removal of a median combined stellar spectrum of KELT-9b. We apply these corrections across all orders for a given exposure. Orders with significant telluric O$_2$ absorption (outlined in Section~\ref{sec:obs} are excluded from the analysis. To detect the shallow planetary transmission spectral signature, we cross correlate the observed spectra against a synthetic template of the planetary atmosphere. 

\subsubsection{Synthetic model spectrum} \label{sec:cc_models}
In this work, we calculate the absorption cross-sections of each species using the open-source and custom opacity calculator \texttt{HELIOS-K}~\citep{2021Grimm}. We assume Voigt line profiles for the absorption lines, 0.258 km s$^{-1}$ spectral resolution at a reference wavelength of 5000\,\AA{}. We adopt other default settings of HELIOS~K, such as the line-wing cutting length, as per~\citet{2021Grimm}. In this work, we explore the line-list from~\cite{2017Kurucz} to calculate the neutral and singly ionized metals: Fe I, Fe II, Ca I, Ca II, Mg I, Mg II, O I, Sc II, Cr II, Ti I, Ti II, TiO and Y II. 

Although we anticipate recovering atomic species only, some molecules were also explored. The line-lists for the molecules investigated are H$_2$O~\citep{Barber2006}, CH$_4$~\citep{Yurchenko2014}, CrH~\citep{Burrows2002}, SiO~\citep{Barton2013}, SiH~\citep{Yurchenko2018}, VO~\citep{Mckemmish2016}, MgH~\citep{Yadin2012}, and TiO~\citep{Mckemmish2019}. The chemical concentrations in the atmosphere are calculated using the open-source code FastChem (\citealt{2018Stock}). For the transmission spectra, we write a script that takes into account the results from \texttt{HELIOS-K} and FastChem, and is based on the simple formalism presented in~\cite{2017Gaidos} and~\cite{2019Bower}. Our model computes the effective tangent height in an atmosphere that was discretised in 200 annuli (\citealt{2021Cabot},~\citealt{2021Bello}). We include in our model the H$^{-}$ bound–free and free–free absorption from~\cite{1988John}. Each transmission spectrum includes one gas species along with H$^{-}$ continuum absorption and scattering by H and H$_2$. 

The models use the planet bulk parameters presented in~\citet{Gaudi_2017}, and assume the atmosphere is isothermal at its equilibrium temperature.

\subsubsection{Cross correlation against spectral residuals} \label{sec:cc_apply}

A forest of metallic absorption lines are present in the optical wavelengths of highly irradiated hot Jupiters. We perform a cross correlation between the observed spectral residuals and the synthetic planetary spectra described above. 

To reduce edge-induced effects in the cross correlation, we apply a 30\% cosine apodisation to the edge of the observed spectrum. The cross correlation is performed using the \texttt{PyAstronomy} package~\citep{pya}. The cross correlation functions from each order are average combined, weighted by their noise, to a master cross correlation function per exposure. We further correct for the Doppler shadow of the planetary transit as per Section~\ref{sec:lc_extract}.

\begin{figure}
    \centering
    \includegraphics[width=0.8\columnwidth]{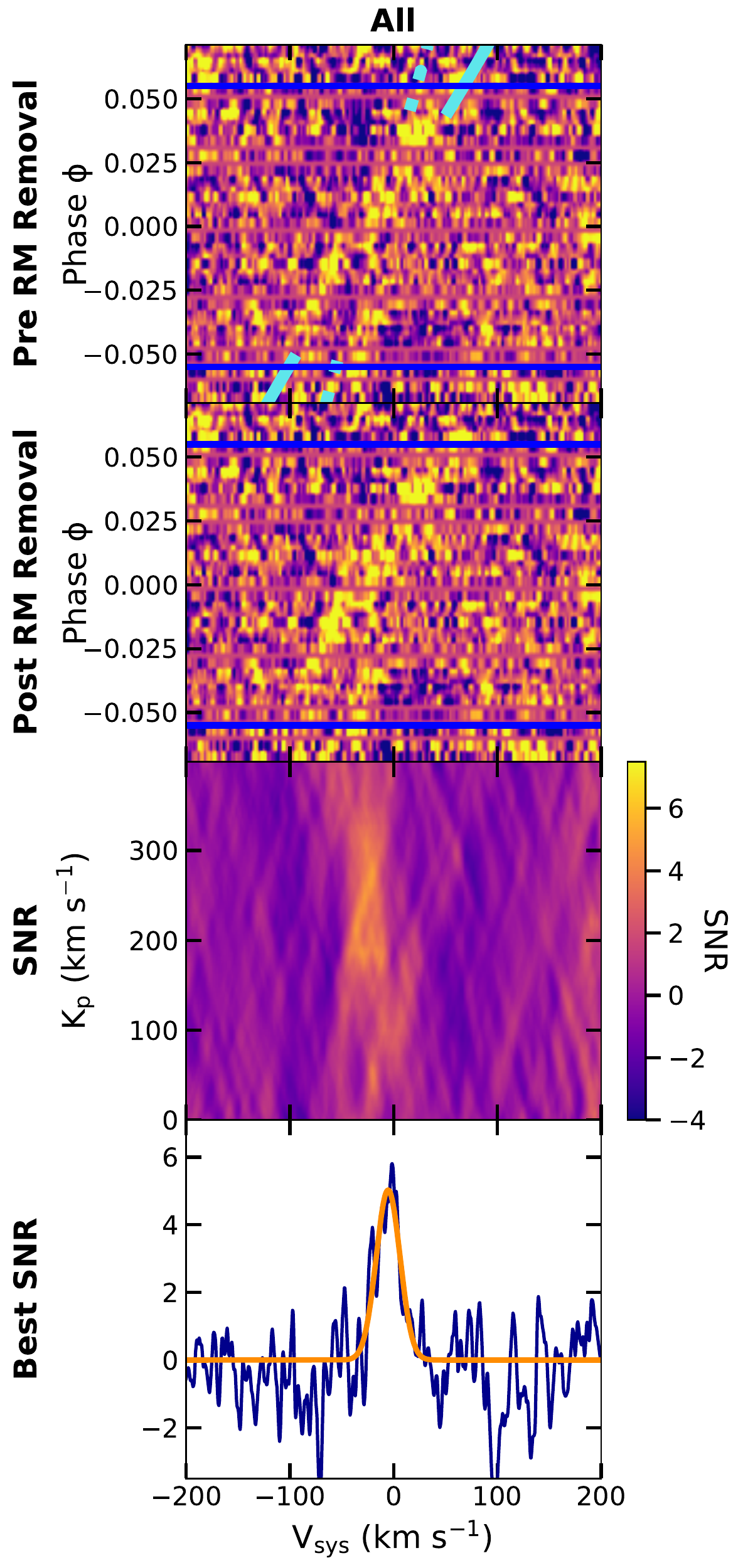}
    \caption{The cross correlation function for the full synthetic absorption spectrum of KELT-9b. Figure description is as per Figure~\ref{fig:h-alpha}.}
    \label{fig:KELT-9b}
\end{figure}

\begin{figure*}
    \centering
    \includegraphics[width=1.7\columnwidth]{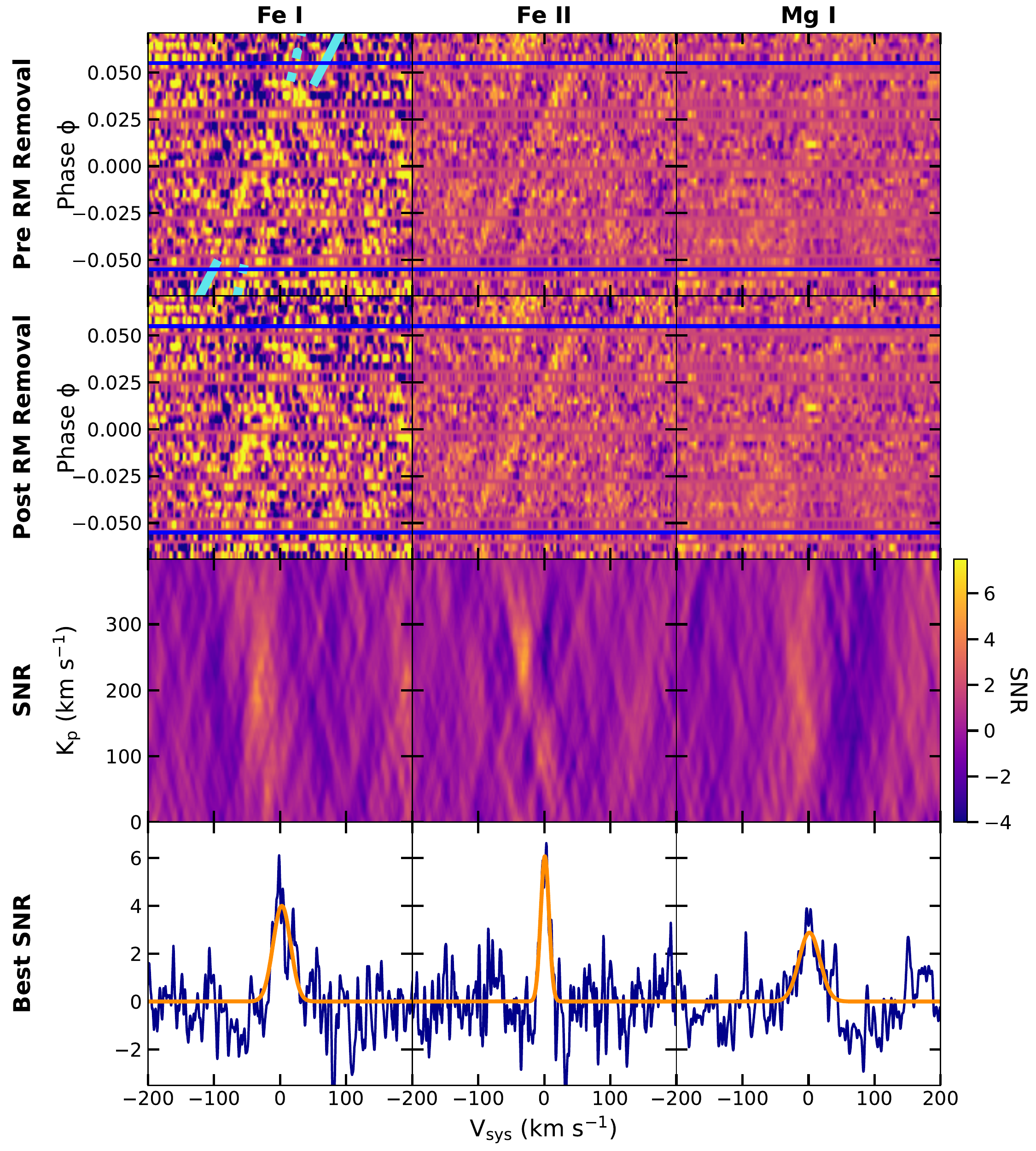}
    \caption{The cross correlation function for Fe I, Fe II, and Mg I. Figure description is as per Figure~\ref{fig:h-alpha}.}
    \label{fig:1415_species_fits}
\end{figure*}

The transmission signature of the atmosphere of KELT-9b was identified in archival TRES transit observations from 2014 and 2015.  Figure~\ref{fig:KELT-9b} shows the transmission spectroscopic signal of KELT-9b from the joint 2014 and 2015 datasets, including the cross correlation function signal-to-noise ratio as a function of the orbital parameters \kp{} and \vsys{}. We report a 6$\upsigma$ detection of the joint 2014 and 2015 datasets. In addition, Fe I, Fe II, and Mg I were individually detected at a significance of 6$\upsigma$, 6$\upsigma$ and 4$\upsigma$ respectively (Figure~\ref{fig:1415_species_fits}). Due to the difference in signal-to-noise between our observations and those from previous literature \citep{Yan_2019, Turner_2020}, we did not recover the transmission signals of Ca I, Ca II, Cr II, Sc II, Ti II, Y II, nor any of the investigated molecules for KELT-9b. The independent atmospheric detection for all epochs are presented in appendix Figure~\ref{appendix:A1}, with independent molecular species analysis for 2014 and 2015 presented in appendix Figure~\ref{fig:2014_species_fits} and appendix Figure~\ref{fig:2015_species_fits} respectively. 

\subsection{Stellar and Planetary Mass Estimates}\label{sec:mass}
Spectroscopic detections of the planetary transit allow us to empirically determine the dynamical masses of both components of the system. In this scenario, the system can be solved as a double lined eclipsing binary, with the stellar radial velocity amplitude $K_\star$ and the planet radial velocity amplitude \kp{} being independently measured. 

The best fit planetary radial velocity amplitude for each species with a strong cross correlation peak is tabulated in Table~\ref{tab:kpvsys}. For each species, we compute the cross correlation function strengths as a function of the planetary radial velocity amplitude \kp{} and the systemic velocity \vsys{} (e.g. Figure~\ref{fig:KELT-9b}). We then fit a 2D Gaussian to the CCF height, \kp{}, \vsys{} surface. As there is significant scatter in the resulting best fit velocity amplitudes, we adopt the standard deviation of the scatter in the solutions for each species as the uncertainty in the subsequent mass calculations.

Rearranging the standard radial velocity equation for an aligned circular orbit \cite[e.g.][]{2018exha.book.....P},
\begin{equation}
    K_\star = \left(\frac{2\text{G}}{P}\right)^{1/3} \frac{M_p}{M_\star + M_p}^{2/3}
\end{equation}
for the stellar mass $M_\star$ we get
\begin{equation}\label{eq:solar_mass_redefined}
    M_\star = K_\star^3\frac{P}{2 \pi \text{G}}\frac{(q+1)^2}{q^3} \,,
\end{equation}
where $q$ is the ratio between the radial velocity amplitude of the star and the planet $q = K_\star/K_p$. Likewise, the planet mass is 
\begin{equation}\label{eq:planet_mass}
    M_p = M_\star q \,.
\end{equation}

Adopting the period $P$ and stellar radial velocity amplitude $K_\star$ from \citet{Gaudi_2017}, along with our measured planetary radial velocity amplitude $K_p=231 \pm 27$ \kms, we get $M_\star=1.91\pm0.68$ \Msol and $M_{p}=2.31\pm0.89$ \Mj. Both our stellar mass and planet mass are consistent with the values reported in \citet{Gaudi_2017} and \citet{Asnodkar_2022} to within $1\upsigma$.

\begin{deluxetable*}{ccc}
\tablewidth{0pc}
\tabletypesize{\scriptsize}
\tablecaption{
        Measurements of orbital and systemic velocities from per-species cross correlations
    \label{tab:kpvsys}
}
\tablehead{
    &
    &
     \\
    \multicolumn{1}{c}{Species}          &
    \multicolumn{1}{c}{Orbital Velocity $K_{p}\,(\kms)$} \tablenotemark{a}             &
    \multicolumn{1}{c}{Systemic velocity $v_{sys}\,(\kms)$}   \tablenotemark{b}         
}
\startdata
    Combined template & $230\pm140$ & $-23\pm18$ \\ 
    H$\upalpha$ & $260\pm110$ & $-24\pm18$ \\ 
    Fe I & $200\pm140$ & $-28\pm18$ \\ 
    Fe II & $260\pm50$ & $-32.6\pm8.3$ \\ 
    Mg I & $200\pm160$ & $-13\pm17$ \\ \hline
    Adopted values for mass calculations & $231\pm27$ & $-25\pm10$ \\
\enddata 
\tablenotetext{a}{
Best fit orbital velocity, quoted uncertainties are the widths of the 2d Gaussian fit along the $K_p$ plane. 
}
\tablenotetext{b}{
Best fit systemic velocity, quoted uncertainties are the widths of the planetary absorption feature. 
}
\end{deluxetable*}

\section{Discussion}\label{sec:results}
In this paper, we sought to characterise the transmission spectrum of the Ultra Hot Jupiter KELT-9b. We report a re-analysis of multi-epoch transits obtained with the TRES facility on the 1.5\,m telescope at FLWO. The observations yielded  detections of excess H$\upalpha$ absorption about the planet, which has been previously cited as a potential indirect tracer for ongoing atmospheric escape. We also report detections of select atomic species in the optical transmission spectrum of KELT-9b. The positive detections of planetary atmospheric features by a meter-class facility opens the possibility of long-term temporal monitoring for highly irradiated planets.

\subsection{H$\alpha$ as a tracer for evaporation and temporal atmospheric variations}
Evaporative processes play a key role in shaping the evolution of close-in exoplanets. Observations of excess planetary absorption of H$\upalpha$ probe the extended neutral hydrogen envelope of planets as they undergo mass-loss. The shape and size of the neutral hydrogen transits can also provide key tests for the interactions between the stellar wind and the escaping hydrogen exosphere. \citet{Owen_2021} notes that the observed Ly$\upalpha$ excess is most dependent on timescale of photoionisation of the neutral hydrogen tail. In strong EUV environments, the neutral hydrogen tail that is optically thick in Ly$\upalpha$ is quickly ionised by the stellar wind, resulting in a reduced transit depth in these wavelengths. 3D simulations~\citep[e.g.][]{Kubyshkina2021} also reinforce the importance of stellar wind interactions on the observed shape and sizes of the escaping neutral hydrogen tails. 

Multi-epoch analysis of the KELT-9b H$\upalpha$ extended atmosphere has the potential of revealing temporal variabilities in the neutral hydrogen tail of the planet. \citet{Cauley2019} reported an asymmetric `W' shaped H$\upalpha$ transit light curve, while \citet{Yan_2018} noted no asymmetric substructure in their transits (although signal-to-noise in their dataset prohibited them to resolve velocity centroids of individual transmission spectra). Our observations agree with \citet{Cauley2019}, and reveal a `W' shaped transit in both our 2014 and 2015 TRES transits (Figure~\ref{fig:h-alpha_lc}). A `W' shape is recovered when constructing a spherical transit model that incorporates the CLV variations as outlined in Section~\ref{sec:centretolimb}, however its contribution appears insufficient at explaining all the observed asymmetries. We offer a tail model as one possible scenario to explain the remaining asymmetries. In this scenario, we treat our H$\upalpha$ light curve as being induced via an occultation of escaping material in the form of a half ellipsoid, with the shorter axis with length approximately that of the radius of the planet $(R_{A} \approx R_p)$, and an elongated axis trailing away from the planet in the form \rb{} with a tilt of $\uptheta$. With these additional free parameters, this model sufficiently explains the remaining asymmetries seen in the light curve, and also offers a better fitting model that can be used to search for any temporal variability in the transit shapes, though none were detected at $>3\sigma$ significance.

To test the robustness of our models, we perform a Bayesian inference criterion (BIC) at each epoch, comparing the difference between the tail vs spherical transit scenarios. We find that the tail model is preferred when each epoch is considered individually, and when the observations are modelled together, with $\Delta$BIC$_{2014}=-70$, $\Delta$BIC$_{2015}=-249$, and $\Delta$BIC$_{\mathrm{combined}}=-11$ respectively.

We note this model departs from standard models of atmospheric escape. Past neutral hydrogen transit models tend to assume the outflowing gas trails behind the planet along its orbital path~\citep[e.g.][]{Owen_2021}. This has been justified by the extended Ly$\upalpha$ transits of GJ 436b \citep{2017A&A...605L...7L}. However, `energetic neutral atoms' have been observed to stream radially away from Solar System planets, and has been proposed as a dominant process shaping the Ly$\upalpha$ absorption of hot Jupiters. \citet{2008Natur.451..970H} and~\citet{2010ApJ...709..670E} proposed that the Ly$\upalpha$ excess of HD209458b can be explained by a radial tail of energetic neutral atoms. The radial neutral hydrogen tail is formed when high velocity protons from the stellar wind exchange electrons with the lower velocity neutral hydrogen escaping atmosphere from the planet. The resulting tail streams away from the planet, as it primarily retains the momentum of the stellar wind. The energetic neutral atoms tail has been observed for Venus, Earth, and Mars~\citep{2011SSRv..162..213F}. In addition,~\citet{Owen_2021} notes that ram pressure from the stellar wind is sufficient to induce a significant radial component to the escaping neutral hydrogen exosphere from the planet.~\citet{Mitani_2022} showed through 2D hydrodynamic simulations that ram pressure from the stellar wind particles is sufficient in producing a tail escaping towards the observer's line of sight. They also note that the H$\upalpha$ transit depth does not strongly depend on the stellar mass loss rate. H$\upalpha$ absorption is dominated by the dense inner region of the exosphere, and is more protected from the stellar wind than other escape tracers, such as Ly$\upalpha$.

Typical signatures of atmospheric evaporation, such as the UV Ly$\upalpha$ line and infrared He I 10830\,\AA{} line, are inaccessible to meter-class facilities, but H$\upalpha$ may be an indicator for mass loss that is accessible to meter-class ground-based observations. A number of other planets have reported H$\upalpha$ excess absorption, including KELT-20b \citep[e.g.][]{2018A&A...616A.151C,2019A&A...628A...9C}. Continued monitoring of these transits can help constrain models of stellar wind interactions with escaping planetary atmospheres. Understanding these interactions are key to properly modelling other observational signatures of evaporating atmospheres.

\begin{figure}
    \centering
    \includegraphics[width=0.9\columnwidth]{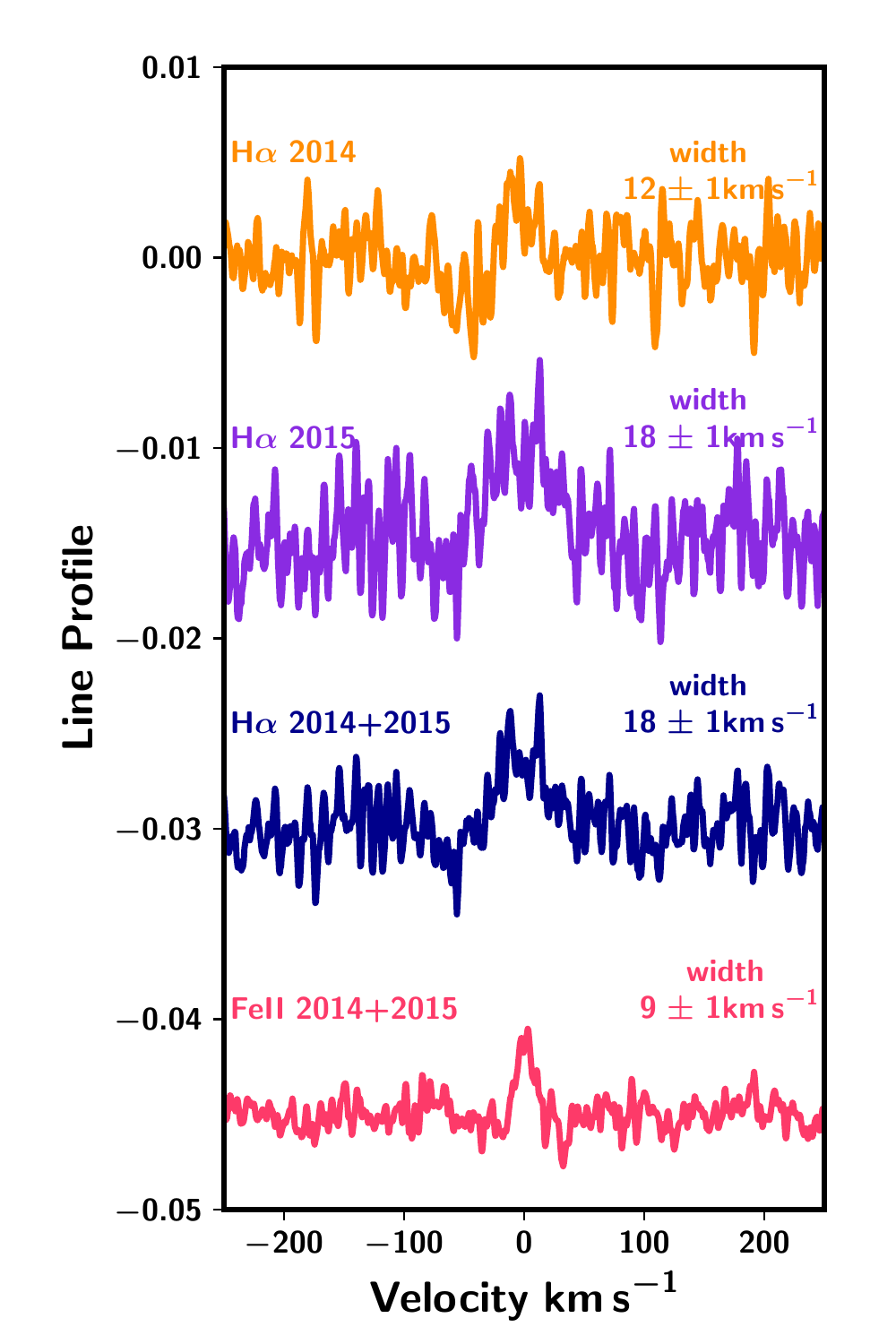}
    \caption{The line profile of the H$\upalpha$ absorption feature, compared to that for Fe II. H$\upalpha$ shows significant line broadening at the $\sim$20 km s$^{-1}$ level in both 2014 and 2015. For comparison, the width of photospheric lines like Fe II are consistent with instrument broadening. Extended broadening of H$\upalpha$ is consistent with the material forming an escaping exosphere about the planet.}
    \label{fig:ha_profile}
\end{figure}

Figure~\ref{fig:ha_profile} shows the observed velocity profiles of the neutral hydrogen in KELT-9b, with the photospheric absorption profile of Fe II being plotted for comparison. The H$\upalpha$ profile is significantly broader than that of the metallic absorption features in the transmission spectrum. We note that no significant blueshift is seen in the H$\upalpha$ line, as would be expected for rapidly escaping gas being accelerated by the stellar wind. This is consistent with the H$\upalpha$ velocity profile from~\citet{Yan_2018}. Models of `energetic neutral atoms' tails of HD 209458b~\citep{2008Natur.451..970H} predict that the Ly$\upalpha$ line profile should be broadened by $\sim$100 \kms{}, and be somewhat blueshifted. If H$\upalpha$ traces escaping neutral hydrogen for KELT-9b, it likely probes a much deeper zone in the exosphere, and thus may not exhibit such a dramatic velocity broadening.

\subsection{A lack of $n=2$ excited neutral hydrogen in the exosphere}
\label{sec:n2}

Significant H$\upalpha$ absorption stemming from the extended tail requires a high number density of neutral hydrogen at the $n=2$ excited state. Maintaining a large population of neutral hydrogen at the $n=2$ excited state is a challenge outside of the atmosphere of the planet. \citet{Christie_2013} demonstrated that neutral hydrogen at the metastable $n=2$ state provides significant absorption only for the optically thick parts of the atmosphere. Extended atmospheres are dominated by ionised hydrogen, which provides little absorption over the H$\upalpha$ line. As is, there are significant issues with the interpretation that the observed `W' shaped H$\upalpha$ transit is induced by an extended $n=2$ tail escaping the planet. For a given population of H$\upalpha$ that might form at the upper boundary of the thermosphere, the sudden decrease in temperature at the thermopause could revert the majority of the population back to the ground state or become ionised.

Additional MHD modelling has also demonstrated that the $n=2$ population is not shaped by the stellar wind in the same manner as Ly$\upalpha$ absorption \citep[e.g.][]{Mitani2022}, and as such may not follow a cometary tail shaped outflow as suggested by our modelling.

We note that our escaping tail toy model for the `W' shaped H$\upalpha$ transit of KELT-9b may not present a physical representation of the shape of the H$\upalpha$ envelope for the planet, but instead propose our model be used a tracer for temporal variability amongst epoch observations that present `W' shaped light curves. Improved centre-to-limb non-LTE modelling of the H$\upalpha$ stellar absorption feature may suggest that H$\upalpha$ transit light curves should be naturally `W' shaped without invoking exotic transit geometries.

\subsection{Detection of metallic transmission spectroscopic signatures from metre-class telescopes}
The work in this paper was achieved using the TRES spectrograph on the 1.5 m FLWO reflector. Our work demonstrates the role that small meter-class ground-based telescopes can potentially provide towards future exoplanet atmosphere characterisation, especially for monitoring the interactions between evaporating atmospheres and the stellar environments they reside in.
In addition to the H$\upalpha$ extended atmosphere of KELT-9b, we also successfully recovered the transmission spectrum from Fe I, Fe II and Mg I at significance of 6$\upsigma$, 6$\upsigma$ and 4$\upsigma$ respectively. Due to the lower signal-to-noise of our observations compared to literature observations of KELT-9b, we report a null-detection of Ca I, Ca II, Cr II, Sc II, Ti II and Y II, previously reported to be present in high resolution transmission spectra of the planet. KELT-9b is the first exoplanet to have atomic Fe I and Fe II directly detected in its atmosphere. These elements, typically found in cloud condensates in cooler atmospheres, are present in their atomic and ionised forms in the highly irradiated upper atmosphere of KELT-9b~\citep{Heng_2016, Stevenson_2016, 2018_Hoeijmakers}. The stronger presence of Fe II vs Fe I is further confirmation on the high temperatures being achieved in the upper regions of the atmosphere~\citep{2018_Hoeijmakers, Hoeijmakers_2019, Pino_2020}. Our detection of Mg I is the third KELT-9b observational data set to achieve this, succeeding~\citet{Cauley2019} and~\citet{Hoeijmakers_2019}. \citet{Huang_2017} proposed magnesium to be an important atmosphere coolant due to its electron impact followed by radiative de-excitation ability, and is potentially an indicator for evaporation~\citep{Bourrier_2015}.

\section{Conclusion}\label{sec:summary}
KELT-9b is the hottest close-in Jovian planet known.  In this paper, we report the re-analysis of archival high resolution spectroscopic transits of KELT-9b from the TRES spectrograph on the 1.5 m reflector at FLWO. These observations, obtained in 2014 and 2015, revealed the extended neutral hydrogen atmosphere of KELT-9b, as well as the presence of atomic species in its upper atmosphere. From these observations we recover signals of Fe I, Fe II, Mg I and H$\upalpha$. Using the velocity of the planetary signal during transit, we estimate a planetary and stellar mass of $2.31\pm0.89$ \Mj and $1.91\pm0.68$ \Msol respectively, all in agreement with previous publications. The H$\upalpha$ absorption exhibits a `W' shaped transit in both the 2014 and 2015 observations, which we model as evaporated material escaping KELT-9b and travelling radially away from the planet towards the direction of the observer. This is unlikely to be the true cause of light curve trajectory, due to the assumed inability of H$\upalpha$ being unable to sustain a significant absorption $n=2$ population. Therefore, we instead use this model as an example on how to monitor temporal variability amongst individual epochs for `W' shaped light curves. Our findings highlight the potential impact of metre class telescopes in exoplanet atmosphere characterisation, and our model presents an alternative for monitoring `W' shaped light curves where external effects, such as those invoked by CLV, are insufficient at explaining all observed asymmetries.

\section*{Acknowledgements}

We would like to acknowledge and pay respect to Australia’s Aboriginal and Torres Strait Islander peoples, who are the traditional custodians of the lands, the waterways and the skies all across Australia. We thank you for sharing and caring for the land on which we are able to learn. In particular, we pay our deepest respects to all Elders, ancestors and descendants of the Giabal, Jarowair, and Yuggera nations, upon which the analysis for this paper was undertaken. We would also like to acknowledge and pay our deepest respects to the Indigenous American Elders, ancestors and descendants who are the traditional custodians of the land upon which the Fred Lawrence Whipple Observatory is situated. We would like to thank the referee for their comments, as they significantly improved the scientific output of this publication. GZ thanks the support of the ARC DECRA program DE210101893. CH thanks the support of the ARC DECRA program DE200101840. We also thank Jessica Mink for running the TRES pipeline and maintaining the TRES archive. We acknowledge Andrew H. Szentgyorgyi, Gabor F\'{u}r\'{e}sz, and John Geary, who played major roles in the development of the TRES instrument.

\facility{TRES}
\software{Adobe Illustrator~\citep{adobeillustrator}, astropy~\citep{2018AJ....156..123A}, batman~\citep{2015batman}, emcee~\citep{emcee}, FastChem~\citep{2018Stock}, HELEOS-K~\citep{2021Grimm}, Matplotlib~\citep{Hunter:2007}, NumPy~\citep{harris2020array}, PyAstronomy~\citep{pya}, SciPy~\citep{2020SciPy-NMeth}, seaborn~\citep{Waskom2021}, Spectrum~\citep{Gray1994}, telfit~\citep{2014AJ....148...53G}}

\clearpage

\bibliographystyle{aj.bst}
\bibliography{kelt-9b.bib}

\appendix
\section{Additional Figures}\label{appendix:A1}

\begin{figure*}[h!]
    \centering
    \includegraphics[width=0.7\columnwidth]{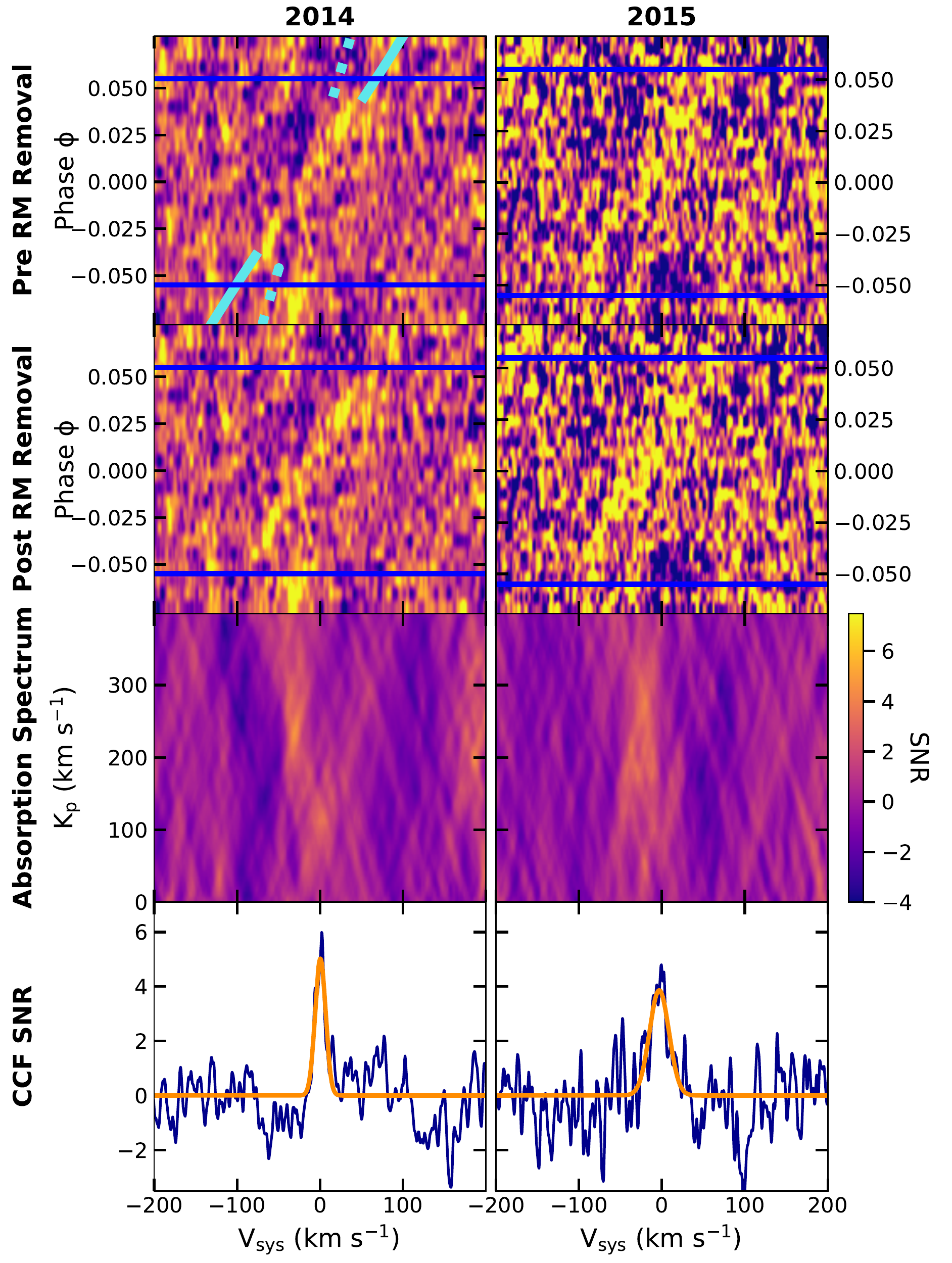}
    \caption{The cross correlation function for the full synthetic absorption spectrum of KELT-9b at the individual 2014 and 2015 epochs. Figure description is as per Figure~\ref{fig:h-alpha}.}
    \label{fig:KELT9b_epchs}
\end{figure*}

\begin{figure*}
    \centering
    \includegraphics[width=\columnwidth]{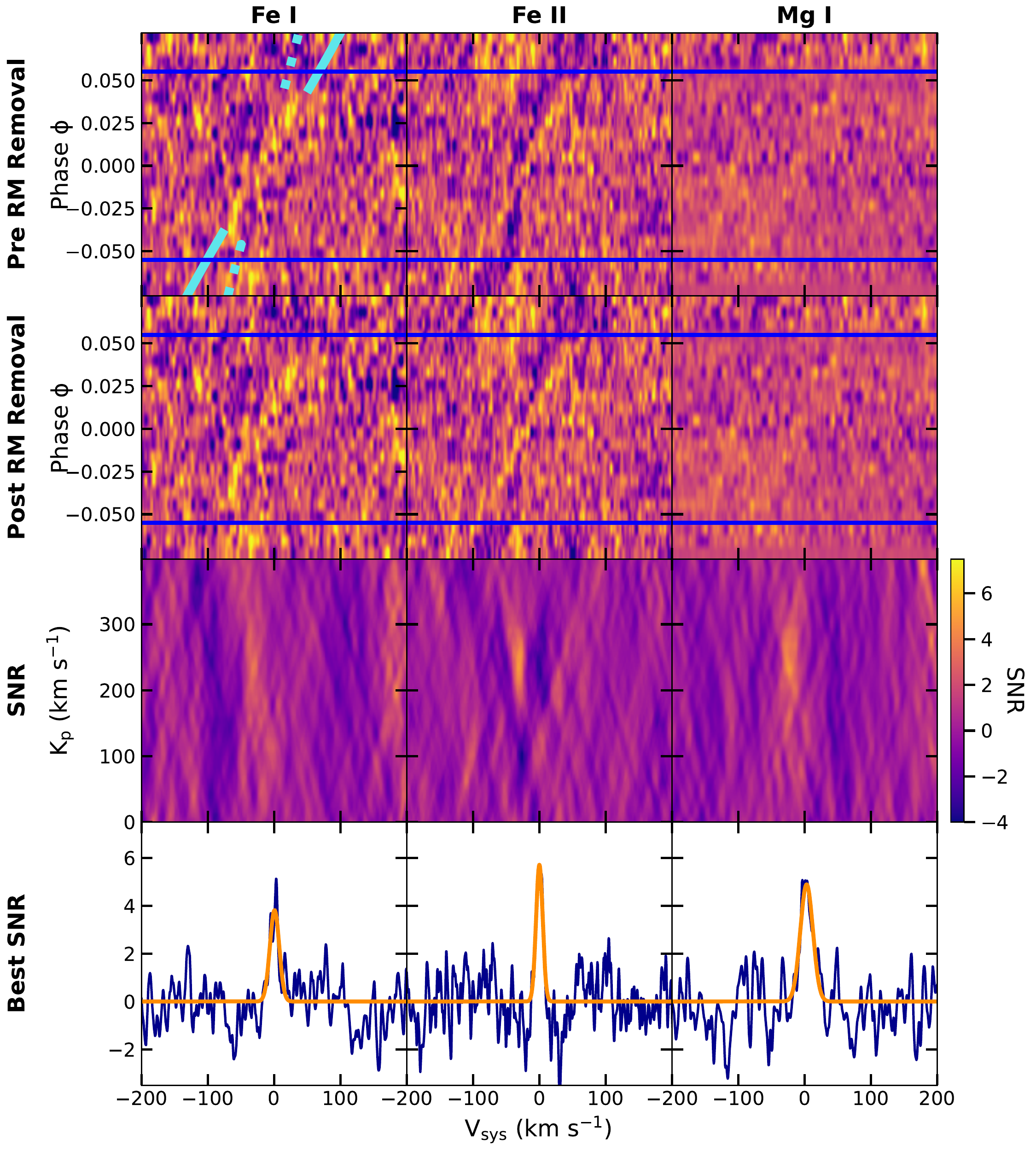}
    \caption{Individual cross correlation analysis for the KELT-9b 2014 epoch using templates for Fe I, Fe II and Mg I. Figure description is as per Figure~\ref{fig:h-alpha}.}
    \label{fig:2014_species_fits}
\end{figure*}

\begin{figure*}
    \centering
    \includegraphics[width=\columnwidth]{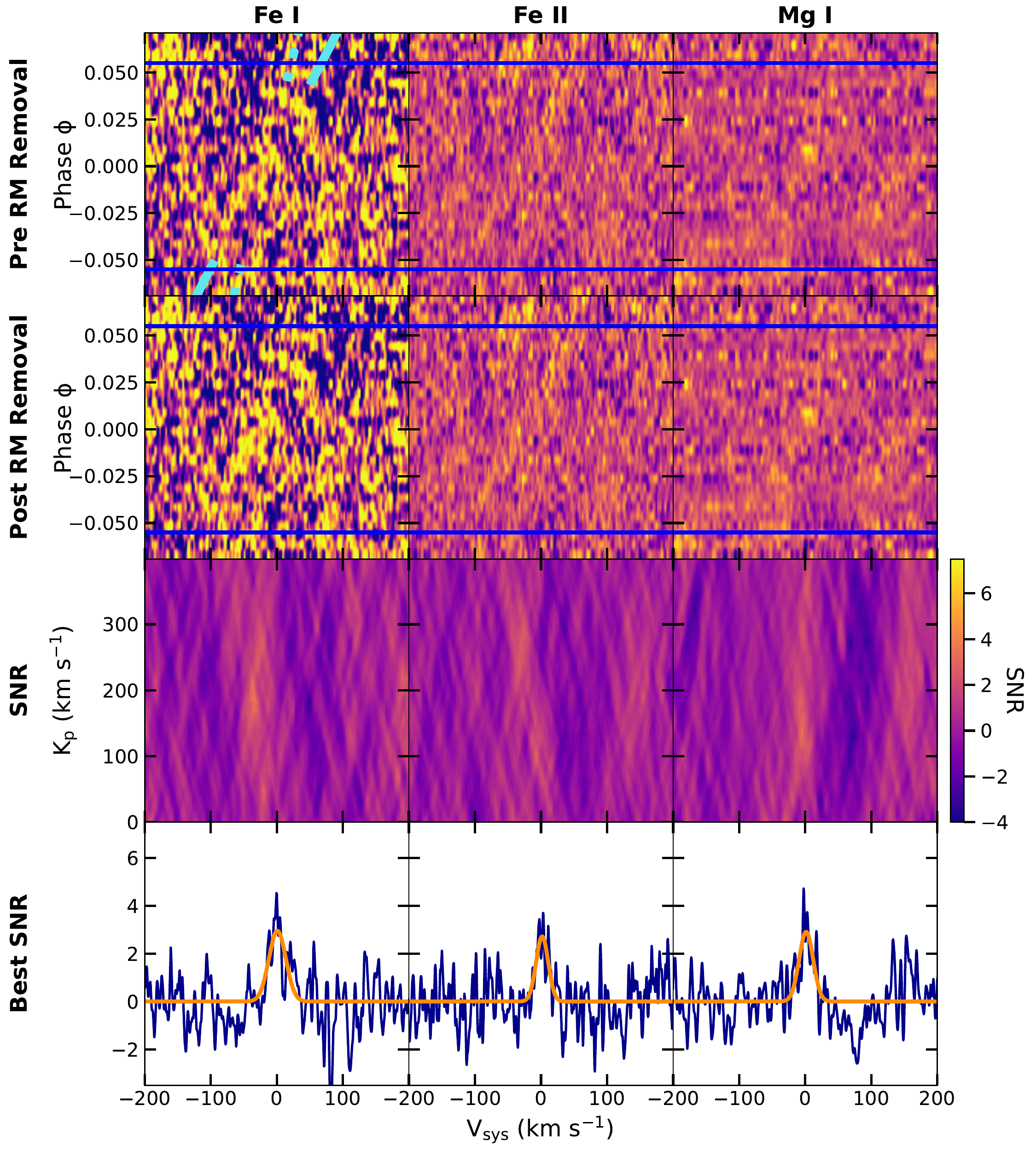}
    \caption{Individual cross correlation analysis for the KELT-9b 2015 epoch using templates for Fe I, Fe II and Mg I. Figure description is as per Figure~\ref{fig:h-alpha}.}
    \label{fig:2015_species_fits}
\end{figure*}


\end{document}